\documentstyle[amssymb,aps,prl]{revtex}

\begin{document}
\title{The phase diagramm of La$_{1-x}$Sr$_x$MnO$_3$ revisited}
\author{M. Paraskevopoulos$^1$, J. Hemberger$^1$, A. Loidl$^1$, A. A. Mukhin$^2$, V.
Yu. Ivanov$^2$ and A. M. Balbashov$^3$}
\address{$^1$Experimentalphysik V, Elektronische Korrelationen und Magnetismus,\\
Institut f\"{u}r Physik, Universit\"{a}t Augsburg, D - 86159\\
Augsburg,~Germany\\
$^2$General Physics Institute of the Russian Acad. Sci.\\
117942 Moscow, Russia\\
$^3$Moscow Power Engineering Institute, 105835 Moscow, Russia}
\date{submitted to PRL Oct.98}
\maketitle

\begin{abstract}
We report on detailed sucseptibility and magnetization studies in La$_{1-x}$%
Sr$_x$MnO$_3$ for $x\leq 0.2$. We arrive at a phase diagram which shows a
ferromagnetic ground state for $0.1\leq x\leq 0.15$ followed by a canted
antiferromagnetic phase at higher temperatures, not in agreement with
present canonical conjectures. The phase diagram indicates the importance of
structural aspects and superexchange interactions for the magnetic phases at
low Sr concentrations.
\end{abstract}

\vspace*{1cm}

It is now half a century ago that Jonker and van Santen \cite{Jonker}
demonstrated that the manganite perovskites La$_{1-x}$A$_x$MnO$_3$, in which
the trivalent La ion is subtituted by a divalent cation, reveal many
interesting and puzzling phenomena. On subtitution with A$^{2+}$, the
antiferromagnetic (AFM) ground state of stochiometric La$^{3+}$Mn$^{3+}$O$%
_3^{2-}$ becomes ferromagnetic (FM) for $x\geq 0.2$ \cite{Jonker,Wollan}. At
that time the FM properties of (La$^{3+}$Mn$^{3+}$O$_3^{2-}$)$_{1-x}$(A$%
^{2+} $Mn$^{4+}$O$_3^{2-}$)$_x$ were explained by a strong positive Mn$^{3+}$%
-Mn$^{4+}$ exchange interaction \cite{Jonker}. In a famous theoretical work
it was de Gennes \cite{deGennes} to establish the close connection of
electrical transport and magnetic properties via an interplay of Mn-O-Mn
superexchange (SE) interactions with Zener's double exchange (DE) \cite
{Zener}. Driven by an increasing concentration of mobile holes, the
insulating (I) and AFM structure passes via a canted AFM (CA) structure to a
purely metallic (M) and FM ground state\cite{deGennes}. Recently an
overwhelming interest in these compounds arose due to the observation of a
negative collossal magnetoresistance (CMR) in a certain composition range
\cite{Chahara}. These CMR effects at the FM phase transition were explained
within extended double exchange models \cite{Furukawa} taking also
Jahn-Teller distortions into account\cite{MillisSolovyev}.

An early structural phase diagram has been presented by Bogush $et$ $al.$
\cite{Bogush}. They showed that on decreasing temperature pure LaMnO$_3$
reveals the sequence rhomboedral (R) to orthorombic (O) and finally to
another orthorombic structure (O\'{}). In the O phase the three octahedral
Mn-O bond lengths are almost equal, while in the O\'{} phase, due to a
long-range Jahn-Teller (JT) distortion, these bond lengths become strongly
anisotropic. On increasing hole doping, and concomitantly with an increasing
concentration of Mn$^{4+}$, the Jahn-Teller distortions become suppressed
and for $0.15<x<$ $0.2$, the undistorted O phase extends to the lowest
temperatures. An electronic phase diagramm for La$_{1-x}$Sr$_x$MnO$_3$ ($%
x\leq 0.5$) has been published by Urushibara $et$ $al.$ \cite{Urushibara}.

That the $x-T$ phase diagram in La$_{1-x}$Sr$_x$MnO$_3$ for low doping
concentrations ($x<0.2$) is even more complex, has been established by
Kawano $et~al.$ \cite{Kaw96}, Yamada $\ et~al.$ \cite{Yam97} and Zhou $%
et~al. $ \cite{zhou97}. In these phase diagrams it has been explicitly
assumed that the regime where the resistivity decreases with decreasing
temperature ($\frac{d\rho }{dT}>$ $0$) is metallic and reveals a simple
ferromagnetic spin arrangement and that the low temperature insulating
ground state ($\frac{d\rho }{dT}~<0$) exhibits a CA structure for $x<0.15$.
Based on systematic heat capacity, magnetic susceptibility, magnetization
and magnetoresistance experiments on single-crystalline La$_{1-x}$Sr$_x$MnO$%
_3$ with concentrations $0\leq x\leq 0.2$ we will show that these phase
diagrams are not correct. Our results reveal that in this doping regime the
Jahn-Teller distortions play a more fundamental role than the ferromagnetic
double-exchange interactions.

Single crystals of La$_{1-x}$Sr$_x$MnO$_3$, with concentrations $0\leq x\leq
0.2$ were grown by the floating zone method with radiation heating in air
atmosphere. X-ray diffraction of crushed single crystals revealed
high-quality single-phase materials. However X-ray topography indicated
twinned crystals. An over-all phase diagram for $x<0.5$ and T $<1000$~K has
been published recently \cite{Mukhin}. To establish a detailed and complete
phase diagram, the magnetic suscebtibility and magnetization was measured
using an ac susceptometer in fields up to 16 T and for temperatures 1.5 K $<$
T$~<$~300 K and a dc SQUID magnetometer for fields up to 7\ T and for
temperatures 1.5~K$~<$~T$~<$~300 K.

In Fig.1 the magnetic ac-susceptibility of La$_{1-x}$Sr$_x$MnO$_3$ is
plotted versus temperature for various doping levels. In the low doping
regime (inset of Fig. 1), a pronounced peak at about 125 K -140~K indicates
the paramagnetic (PM) to CA transition. The transition temperature \ $%
T_{CA}\ $ significantly shifts towards low temperatures as the Sr
concentration is increased. A second anomaly at about 180 K probably
indicates a further structural anomaly. For the samples with $x=0.1$ and $%
x=0.125$, one can identify three transitions as the temperature is lowered.
For this concentrations a small anomaly, which indicates a structural
transition, precedes the low-temperature magnetic transition ($x=0.1$: $%
T_{CA}=150$ K, $T_{\text{O\'{}O}}=115$ K, $T_C=105$ K; $x=0.125$: $%
T_{CA}=180 $ K, $T_{\text{O\'{}O}}=155$ K, $T_C=140$ K). For $x=0.15$ this
sequence of structural and magnetic transitions appears in a narrow
temperature range. It will be the aim of this paper to provide experimental
evidence for the nature of this magnetic and structural transitions. For $%
x=0.175$ we detected a strong increase of $\chi _{ac}$ at $T_C=285$ K
followed by a significant decrease on passing the rhomboedral to orthorombic
phase transition at $T_{RO}=185$ K. From the plot of the inverse
susceptibility we denoted one further phase transition at higher
temperatures ($x=0.1$ : $T_{\text{OO\'{}}}=310$ K; $x=0.125$ : $T_{\text{%
OO\'{}}}=270$ K). These transitions are also clearly indicated by a
significant increase of the resistivity on passing from O$\rightarrow $O\'{}
\cite{Mukhin}. At this point we clearly want to state that a intepretation
of the structural anomalies was only possible taking the published neutron
scattering results into account \cite{Kaw96,Yam97,Argyriou,Pinsard}

In the next step we investigated the magnetic ground state of La$_{1-x}$Sr$%
_x $MnO$_3$. Fig. 2 shows the magnetization at 10 K for various compositions
($x\leq $ $0.2$) as a function of the magnetic field. Fig. 2 demonstrates
that the samples with $x<0.1$ are in a CA phase, which we conclude from the
fact that there exists a spontaneous magnetization $M_S$ which is followed
from a linear increase of the magnetization $M$ as the magnetic field is
further raised. It is remarkable that there is no sign of saturation even at
an applied field of 14~T. This behavior corresponds to the predictions of de
Gennes \cite{deGennes} for the manganites in the canted phase. The basic
effect is that the external magnetic field enforces a continuous reduction
of the canting angle. In addition, the initial value of the spontaneous FM
magnetization increases with $x$, a fact that directly demonstrates the
reduction of the canting angle on hole doping. The inset a) of Fig. 2 shows
FM hysteresis loops for this doping regime. Both the remanent magnetization
and the remanent field increase on increasing hole doping. For higher doping
rates ($x\geq 0.1$) we find the typical characteristics of an FM state with
the magnetization reaching saturation within 2 T. There is a complete
parallel allignment of the Mn$^{3+}$/ Mn$^{4+}$ spins as can be seen from
the value of the saturated magnetization which aproaches the classical
spin-only value of 3.9 $\mu _B$ / Mn ion (inset b) of Fig. 2). We would like
to state that for concentrations $0.1\leq x\leq 0.2$ the FM saturation looks
very similar within the experimental uncertainities. Of course, we can not
exclude a small canting angle ($\theta \lesssim 10{{}^{\circ }}$) but there
is definitely a significant change in the magnetic ground state between $%
x=0.075$ and $x=0.1$. For all further discussions we assume a weak
ferromagnetic ground state (large canting angles) for $x\leq 0.075$ and
strong ferromagnetism (low or zero canting angle) for $x\geq 0.1$.

In the following we concentrate on the sample La$_{0.9}$Sr$_{0.1}$MnO$_3$,
although the same conclusions can also be drawn for $x=0.125$. In order to
understand the evolution of the different magnetic states as a function of
temperature and magnetic field, we have measured the magnetization up to 14
T at various temperatures. Fig. 3 shows some representative results. For T $%
> $ $T_{CA}=150$ K the magnetization curves reveal the signature of a
typical paramagnet. As the temperature is lowered below $T_{CA}$, we observe
the evolution of a spontaneous magnetization $M_S$, which growths upon
further cooling. As the external magnetic field is raised there is at first
no obvious sign of saturation. The magnetization seems to behave like a
canted AFM. At a given value of the applied field, characteristic for each
temperature, jumps in the mangetization occur inducing a higher
magnetization. The first loop at lower field signals a field-induced
structural transition \cite{Assamitsu}, from the JT-ordered phase O\'{} to
the orthorombic phase O, accompanied by a step-like reduction of the canting
angle which yields a higher magnetization. At the subsequent second
hysteresis loop the sample undergoes a further transition into a FM state
reaching the full magnetization. With decreasing temperature this anomalies
are shifted to lower temperatures and finally disappear below $T_C$ $=105$ K.

Similar observations, as those presented in Fig. 3 have also been derived
for crystals with a Sr concentration $x=0.125$, although in this case the
two subsequent hysteresis loops are not so nicely separeted and it seems
that the structural and magnetic phase transitions are closer coupled in
temperature. Finally, for $x=0.15$ we were not able to detect any hysteresis
loops, neither in magnetization nor in magnetoresistance measurements.
However, it seems clear that ferromagnetism is established below $T_C\simeq
200$ K, while for temperatures $200$ K $<$ T $<240$ K the magnetization is
not saturated in fields up to 14 T. But the $M(H)$ curves for magnetic
fields $H>0.5$ T always reveal a continuous curvature and never show a
constant slope as observed for $x=0.1$ at T $=145$ K, characteristic for a
canted structure. We are not able to detect clear indications of field
induced structural phase transitions. It is possible that in this
concentration regime all transitions appear in a narrow temperature range
but we favour an explanation in terms of a phase separation.

Based upon these experimental results we tried to construct a structural,
magnetic and electronic phase diagram which is compatible with the published
experimental results. The phase diagram is simple at low strontium
concentrations ($x<0.1$). The transition into the CA ground state is clearly
documented in the inset of Fig. 1 and in Fig. 2. As shown in the inset of
Fig. 1, there exist slight anomalies in the susceptibility just below 200 K.
It is unclear if they signal some kind of structural phase transition.
Recently, a significant anomaly in the temperature dependence of the orbital
order parameter has been observed in pure LaMnO$_3$ just above $T_N$ \cite
{Murakami}. In this system, driven by the long-range JT distortion orbital
order is established at 800 K.

The phase diagram becomes much more complicated at higher concentrations.
Our results provide clear experimental evidence that for $0.1\leq x\leq 0.15$
the ground state is, within experimental accurancy, ferromagnetic and is
followed for $x=0.1$ and $x=0.125$ by a canted spin state at elevated
temperatures. And indeed, using neutron diffraction techniques, Argyiou $et$
$al.$ \cite{Argyriou} have observed the temperature dependence of the
canting angle $\Theta $ for $x=0.125$ and observed a continuous decrease
between 220 K to 160 K and a lock-in like phenomenon of the canting angle at
approximately 20${{}^{\circ }}$ close to 160 K, in close agreement with our
experimental results. A very weak but finite AFM Bragg reflection has also
been reported by Kawano $et$ $al.$ \cite{Kaw96} for the same Sr
concentration which again indicates an almost vanishing canting angle.

This observation of the sequence PM$\rightarrow $CA$\rightarrow $FM is in
clear contradiction to all experimental phase diagrams published so far and
of course calls for a different physical interpretation of the coupling of
electronic and magnetic phenomena. It is clear now that the sequence
metallic ($\frac{d\rho }{dT}>$ $0$)$\rightarrow $insulating ($\frac{d\rho }{%
dT}<$ $0$) can not be explained within a simple double-exchange picture wich
rests upon the sequence FM$\rightarrow $CA. We see no possibility to explain
these results by an electronic phase separation as has been proposed
theoretically \cite{Yunoki} and has been observed experimentally in the Ca
doped system \cite{Allodi}. In pure AFM clusters the hysteresis loops would
shift to higher fields as the temperature is lowered, opposite to what is
observed experimentally (see Fig. 3).

For a correct interpretation of our results it is necessary to take into
account the results of recent neutron diffraction investigations \cite
{Kaw96,Pinsard} which reveal that close to $x=0.125$ the pseudocubic
high-temperature orthorhombic phase O, is followed by the JT distorted O\'{}
phase and finally transforms again into a pseudocubic O\'{}\'{} phase which
reveals superstructure reflections due to charge order \cite{Yam97}. This
O\'{}/O\'{}\'{} transition is clearly indicated in the resistivity results
\cite{Mukhin} but also in the susceptibility data (see Fig. 1). The
structural OO\'{} phase transitions which appear close to room temperature
for $0.1\leq x\leq 0.125$ can be determined from the inverse susceptibility $%
vs.$ temperature (not shown here).

All the observed anomalies are indicated in the $x-T$ phase diagram which is
shown in Fig. 4. At low concentrations ($x<0.1$) we find an insulating
orthorhombic phase O\'{} which reveals canted antiferromagnetism at low
temperatures. In a narrow temperature range at elevated temperatures these
phase extends up to $x=0.15$. Again the PM insulator is followed by a
insulating CA phase. The ''metallic behavior'' ($\frac{d\rho }{dT}<$ $0$) in
this regime obviously results from the freezing-out of spin-disorder. For $%
0.1\leq x\leq 0.15$ the ground state is a FM insulator with a pseudocubic O
phase and charge (Mn$^{3+}$/Mn$^{4+}$) order, further denoted as O\'{}\'{}
\cite{Pinsard}. At the O\'{}O\'{}\'{} transition, which is followed by the
FM transition, the resistivity steeply increases. For $x>0.17$ the ground
state is a FM metal revealing an O structure for $x<0.2$ and a rhombohedral
structure for $x>0.2$. The phase diagram around strontium concentrations
0.15 is unclear. Here the different magnetic and structural phase boundaries
meet and phase separation may be a natural explanation of the experimental
results.

Here we recall the most important result of this study. Close to $x=0.125$
we provide experimental evidence for the sequence: O/PM$\rightarrow $O\'{}/PM%
$\rightarrow $O\'{}/CA$\rightarrow $O\'{}\'{}/CA$\rightarrow $O\'{}\'{}/FM.
From our magnetoresistance measurements, which will be published in a
forthcoming paper we found a positive MR in the O\'{}\'{}/CA phase, while
large and enhanced negative MR effects appear close to the O\'{}/PM to
O\'{}/CA phase boundary. A positive MR has also been reported by Senis $et$ $%
al.$\cite{Senis}.

A simple explanation of the observed structural and magnetic phase
diagram can be given by taking SE interactions into account. For
$x=0$, in the pure compound the MnO$_6$ octahedra reveal a cooperative
rotation about the orthorhombic b-axis. This rotation produces
anisotropic Mn-O-Mn bonds, which in turn give FM exchange interactions
within the a-b planes and AFM ineractions along the c-axis and yield
the A-type AFM structure of LaMnO$_3$ \cite{zhou97,Toepfer}. In the
non-JT distorted O phase the SE is isotropic, yielding
three-dimensional FM interactions between Mn$^{3+}$ and Mn$^{3+}$ as
well as Mn$^{3+}$ and Mn$^{4+}$. Hence the low-temperature structures
of O and O\'{}\'{} which both are pseudocubic, reveal FM order. Within
the O\'{} phase the decrease of the canting angle may well be
determined by DE interactions and DE certainly plays an important role
for $x>0.175$, but the ferromagnetism of the O\'{}\'{} phase
definetely is determined by structural properties, namely the
isotropic superexchange interactions of the pseudocubic perovskite
structure.

This work has in part be supported by the BMBF under the contract number
13N6917.

\clearpage

FIGURE\ CAPTIONS:

Fig. 1: AC susceptibility $vs.$ temperature in La$_{1-x}$Sr$_x$MnO$_3$ for
concentrations $x=0.1,0.125$ and $0.15$. The inset shows the
susceptibilities for $x=0,0.05$ and $0.075$ on a semi-logarithmic scale.

Fig. 2: Magnetization $vs.$ magnetic field for La$_{1-x}$Sr$_x$MnO$_3$ for
concentrations $x=0,0.05,0.075,0.1$ and $0.2$ at 10 K. The inset a) shows an
enlarged view of the FM hysteresis loops at small fields. Inset b) shows the
spontaneous magnetization $M_S$ as a funtion of $x.$

Fig. 3: Magnetization $vs.$ field as observed in La$_{0.9}$Sr$_{0.1}$MnO$_3$
at different temperatures between room temperature and 10 K.

Fig. 4: $x-T$ phase diagram of La$_{1-x}$Sr$_x$MnO$_3$. The structural (O,
O\'{}, O\'{}\'{}, R), magnetic (PM, CA, FM) and electronic (M, I) phases are
indicated. Open symbols (dashed lines) denote structural phase boundaries,
solid symbols (full lines) denote magnetic phase boundaries.

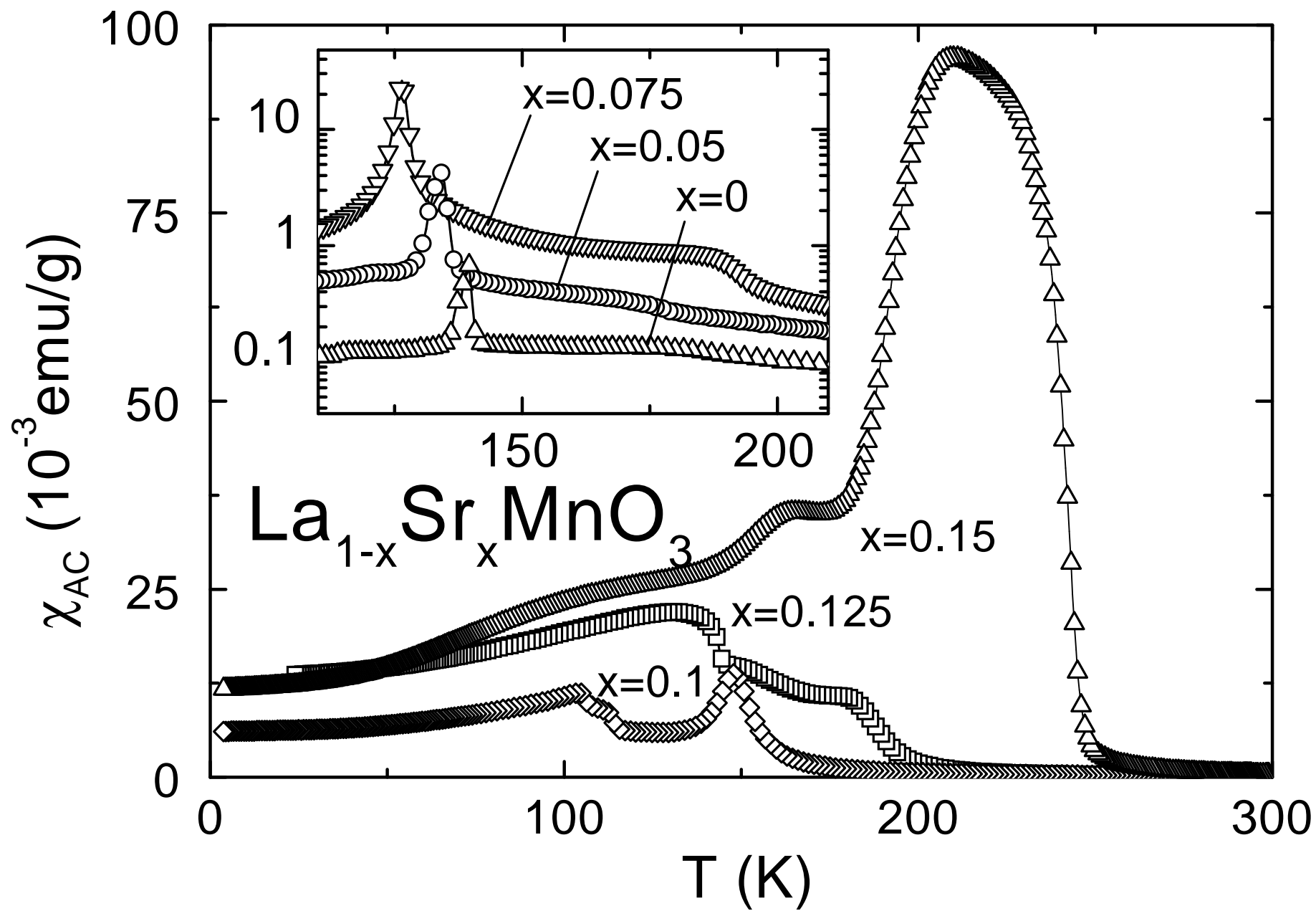
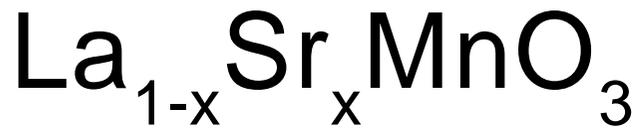
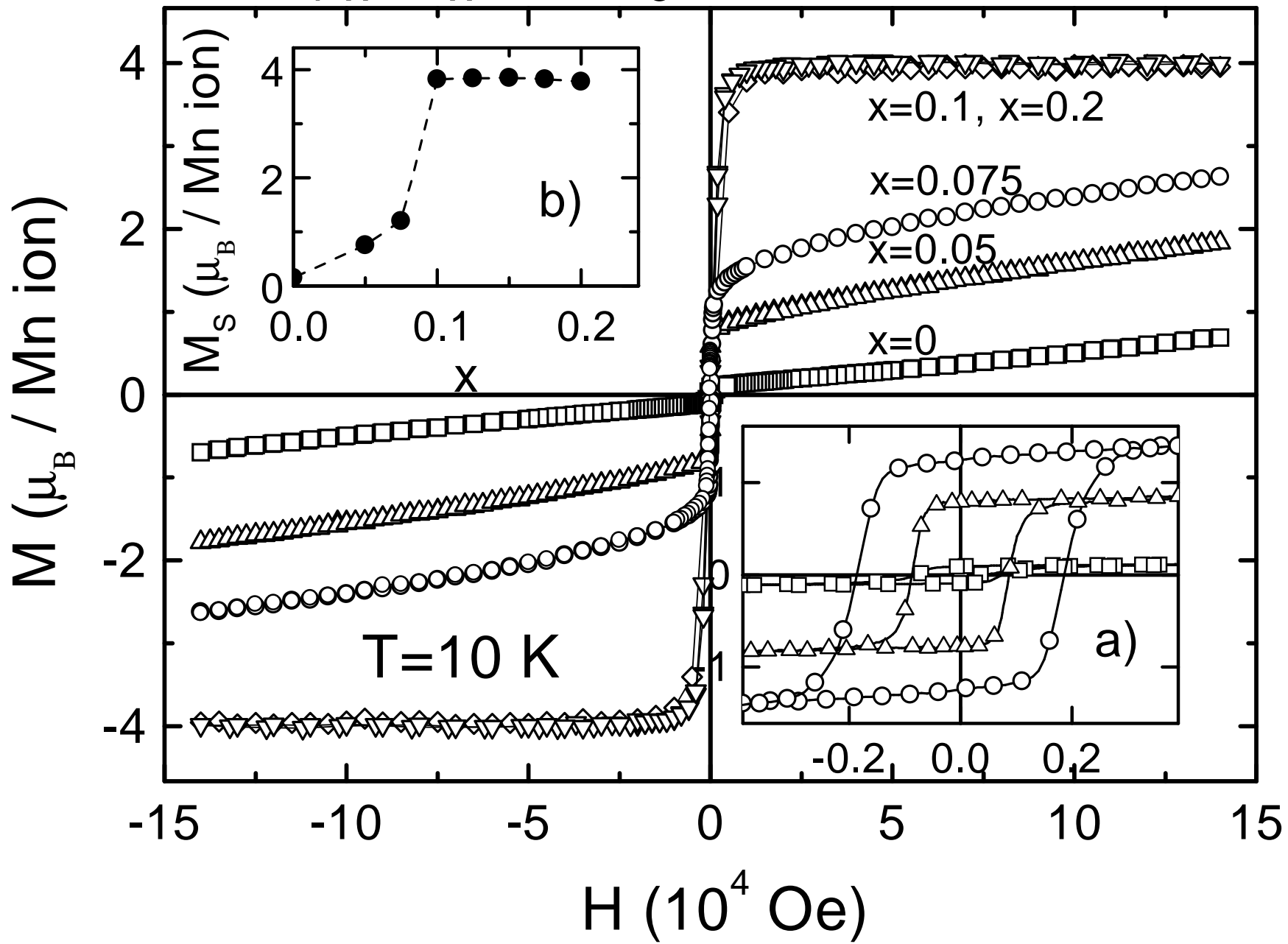
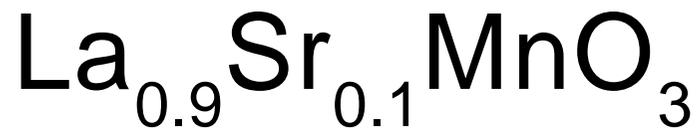
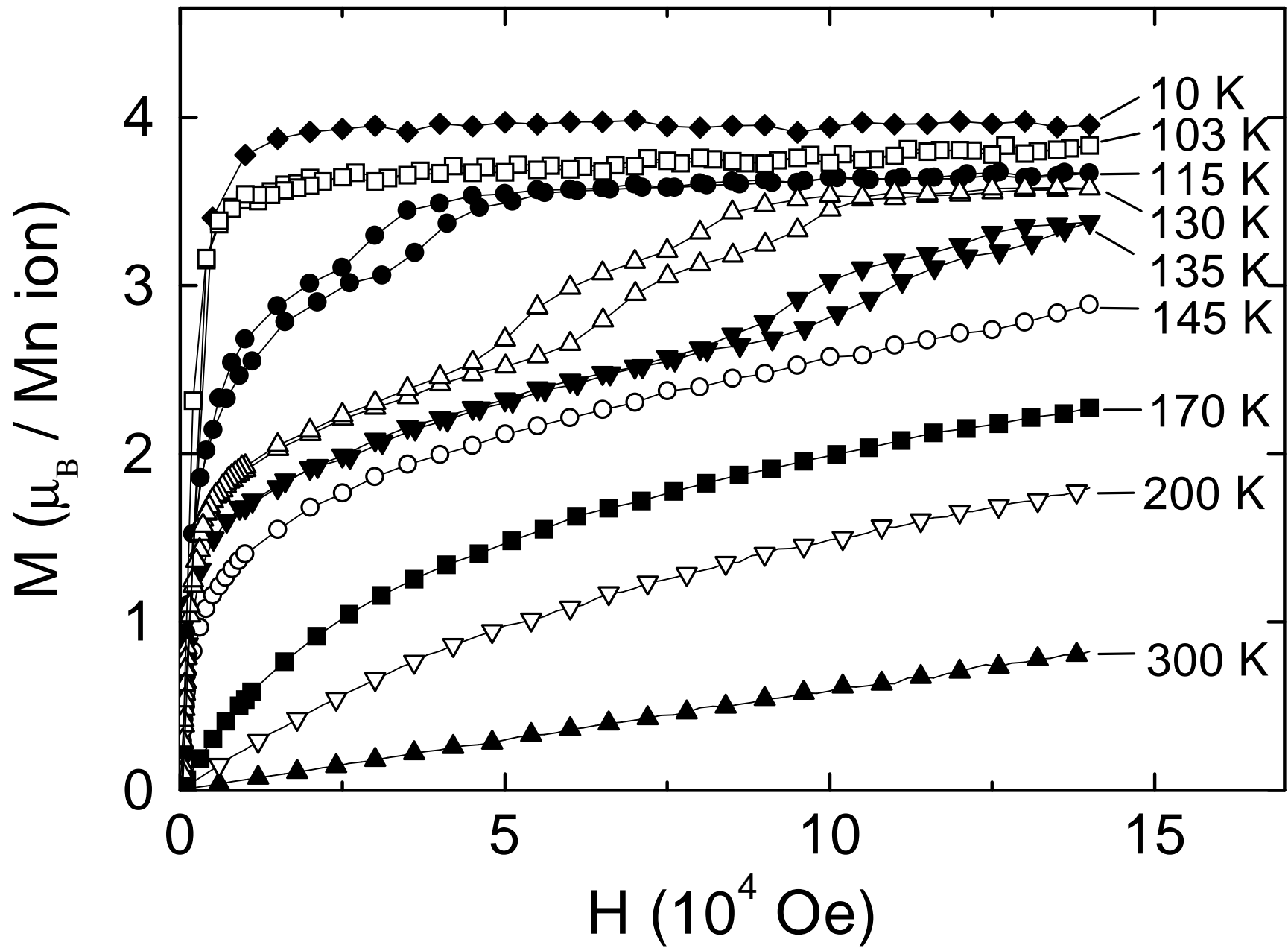
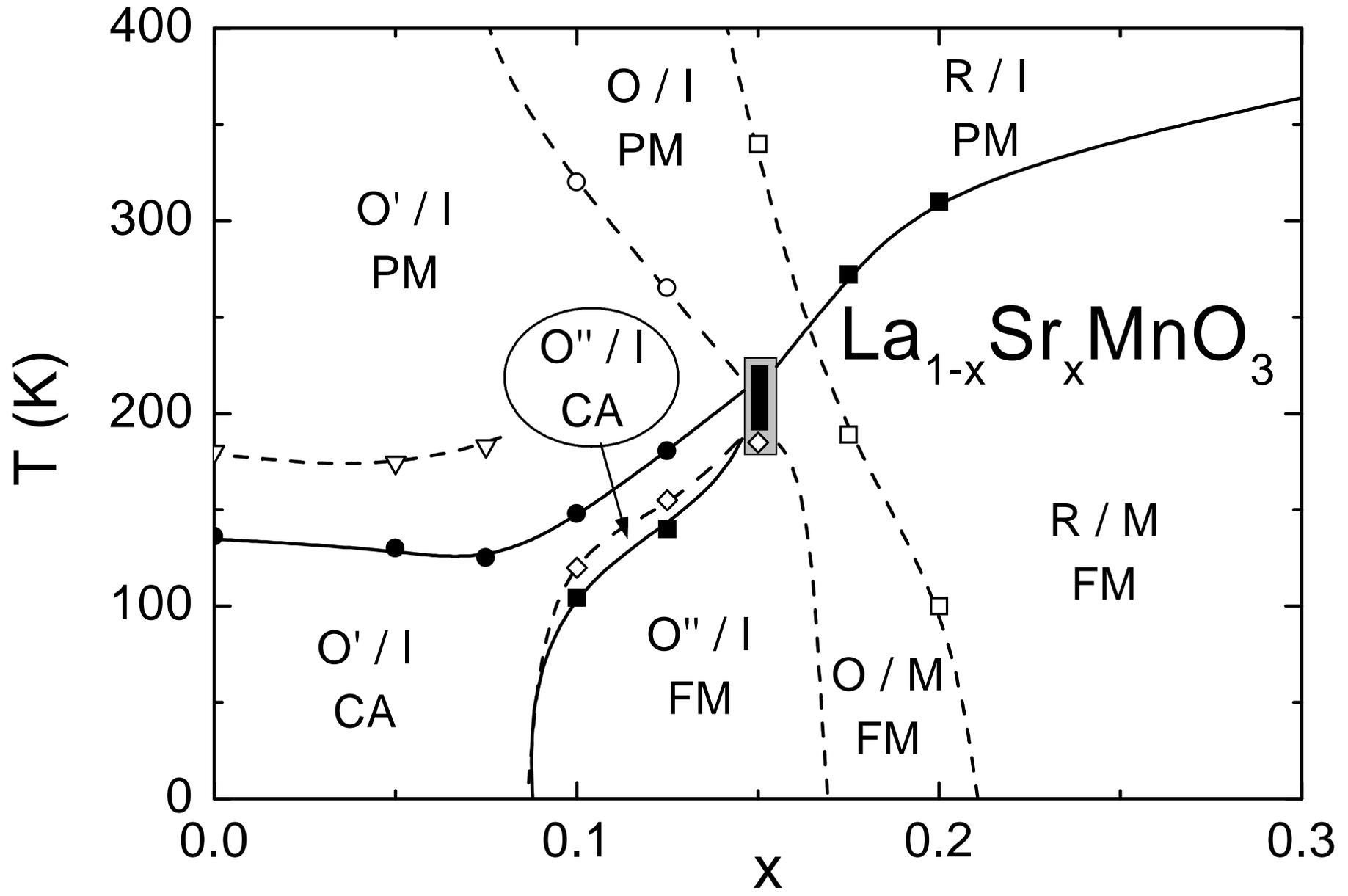
\end{document}